# Compositional dependence of epitaxial Ti$_{n+1}$SiC$_n$ MAX-phase thin films grown from a Ti$_3$SiC$_2$ compound target

Martin Magnuson[a)] Lina Tengdelius, Grzegorz Greczynski, Fredrik Eriksson, Jens Jensen, Jun Lu, Mattias Samuelsson, Per Eklund, Lars Hultman, and Hans Högberg

Department of Physics, Chemistry, and Biology (IFM), Linköping University, SE-581 83, Linköping, Sweden

[a)] Electronic mail: martin.magnuson@liu.se

## Abstract

We investigate sputtering of a Ti$_3$SiC$_2$ compound target at temperatures ranging from RT (no applied external heating) to 970 °C as well as the influence of the sputtering power at 850 °C for the deposition of Ti$_3$SiC$_2$ films on Al$_2$O$_3$(0001) substrates. Elemental composition obtained from time-of-flight energy elastic recoil detection analysis shows an excess of carbon in all films, which is explained by differences in angular distribution between C, Si and Ti, where C scatters the least during sputtering. The oxygen content is 2.6 at.% in the film deposited at RT and decreases with increasing deposition temperature, showing that higher temperatures favor high purity films. Chemical bonding analysis by X-ray photoelectron spectroscopy shows C-Ti and Si-C bonding in the Ti$_3$SiC$_2$ films and Si-Si bonding in the Ti$_3$SiC$_2$ compound target. X-ray diffraction reveals that the phases Ti$_3$SiC$_2$, Ti$_4$SiC$_3$, and Ti$_7$Si$_2$C$_5$ can be deposited from a Ti$_3$SiC$_2$ compound target at substrate temperatures above 850 °C and with growth of TiC and the Nowotny phase Ti$_5$Si$_3$C$_x$ at lower temperatures. High-resolution scanning transmission electron microscopy shows epitaxial growth of Ti$_3$SiC$_2$, Ti$_4$SiC$_3$, and Ti$_7$Si$_2$C$_5$ on TiC at 970 °C. Four-point probe resistivity measurements give values in the range ~120 to ~450 μΩcm and with the lowest values obtained for films containing Ti$_3$SiC$_2$, Ti$_4$SiC$_3$, and Ti$_7$Si$_2$C$_5$.

## I. INTRODUCTION

The class of inherently layered carbide and nitride materials known as $M_{n+1}AX_n$ (n=1 to 3) phases[1-6] have attracted both scientific research and industrial interest because of their unique combination of metallic and ceramic properties. The archetype of this class is $Ti_3SiC_2$,[7,1] extensively studied as bulk and thin films.[8,9] For $Ti_3SiC_2$ thin films, early work focused on chemical vapor deposition as reported by Nickl *et al.* in 1972[10] with several works during the 1980s and 90s.[11-13] While CVD has continued to be studied for MAX phases,[14,15] physical vapor deposition and especially sputter-deposition is much more established today. There are two main approaches: magnetron sputtering from a $Ti_3SiC_2$ compound or other composite target, and growth from elemental targets (Ti, Si, and graphite, or in some of the early work together with $C_{60}$ evaporation).[16] Epitaxial growth of $Ti_3SiC_2$ on MgO(111) substrates was demonstrated at substrate temperatures in the range 900-1050 °C. For fundamental research and exploration of the Ti-Si-C system, sputtering from individual sources became the preferred synthesis route.[17,18] In 2002, the $Ti_4SiC_3$ phase had been theoretically predicted,[19] and sputtering methodology enabled the discovery of the $Ti_4SiC_3$ phase as well as two intergrown phases in the form of $Ti_5Si_2C_3$ and $Ti_7Si_2C_5$, with alternating "211" and "312" or "312" and "413" layers, respectively[17]. In 2011, Scabarozi *et al.* showed that it is possible to deposit almost phase pure and epitaxial $Ti_7Si_2C_5$ by reactive sputtering,[20] and recently the $Ti_4SiC_3$ phase has been demonstrated in bulk.[21-23]

For industrial applications, growth from elemental sputter sources poses limitations for a simple, repeatable and scalable process, shifting the attention back to compound targets. Depositions from a $Ti_3SiC_2$ compound target using substrate temperatures ≤ 300°C[24] typically results in growth of nanocomposite thin films of nanocrystalline (nc) – TiC/amorphous (a)-SiC. Another study by Eklund *et al.*[25] at higher deposition temperatures showed that epitaxial $Ti_3SiC_2$ films containing $Ti_4SiC_3$ and $Ti_7Si_2C_5$ phases could be co-sputtered from a $Ti_3SiC_2$ compound target together with a Ti target on a TiC(111) seed layer on $Al_2O_3$(0001) substrates at 850 °C.

Balzer and Fenker[26] applied high power impulse magnetron sputtering (HiPIMS) to deposit films of 3Ti:Si:2C stoichiometry from a $Ti_3SiC_2$ target without substrate heating. Growth from HiPIMS was also used by Alami *et al.*[27] at a substrate temperature of 680 °C, which resulted in the deposition of crystalline TiC or a phase mixture of the Nowotny $Ti_5Si_3C_x$-phase and TiC depending on the inclination angle of the sputtered material. The Nowotny phase $Ti_5Si_3C_x$ can be considered as a solid solution of carbon in the silicide $Ti_5Si_3$.[28,29]

The results presented above show that growth of $Ti_3SiC_2$ from a $Ti_3SiC_2$ compound target is not straightforward, but highly dependent on the choice of deposition parameters. The coating composition depends on gas-phase scattering processes and differences in angular and energy distributions of sputtered atoms, which is to be expected given the difference in masses between Ti, Si, and C.[25] In contrast, backscattering of Ar neutrals is not an operative effect, as it is for heavy elements, see *e.g.*,[30-32] for sputtering of WTi. For Ti-B films, Neidhardt *et al.*[33] concluded that the stoichiometry of the Ti-B films are highly dependent on the pressure-distance product.

In the present study, we apply a target-to-substrate distance of 7 cm and with the sputter target directly facing the substrate surface, substrate temperatures up to 1000 °C, and an argon pressure of 4 mTorr to deposit $Ti_3SiC_2$ films on $Al_2O_3(0001)$ substrates from a $Ti_3SiC_2$ compound target. Films consisting of $Ti_3SiC_2$, $Ti_4SiC_3$, and $Ti_7Si_2C_5$ MAX-phases were grown.

## II. EXPERIMENTAL CONDITIONS

### A. Deposition

The films were deposited by direct-current magnetron sputtering from a 3-inch circular $Ti_3SiC_2$ compound target from Kanthal, Sweden, in a ultra-high vacuum (UHV) system (base pressure of $\sim 10^{-6}$ Pa) with the substrate positioned directly in line-of-sight above the magnetron at a distance of 7 cm. In all depositions, the argon (99.9997%) pressure was kept constant at 0.53 Pa and growth was carried out at a floating potential corresponding to $\sim$ -25V. To minimize the possibility of crack formation in the target the sputtering power was ramped to the desired power during at least 240 s. To avoid stray coating on the substrates, we used a shutter that completely covered the magnetron. We investigated the growth conditions at substrate temperatures ranging from no applied external heating (RT) to 970 °C, the highest possible temperature for the applied substrate heater, using a sputtering power of 300 W that corresponds to a sputtering current of ~660 mA as seen from a discharge voltage of ~450 V and for a deposition time of 300 s. The film thickness determined at a sputtering power of 300 W was 200 nm corresponding to a growth rate of 40 nm/min at a deposition time of 300 s (5 minutes). At 850 °C, we studied the influence of sputtering power by applying 50, 100, 150, 200, and 250 W to the target, which resulted in sputtering currents of ~130, ~240, ~350, ~450, and ~560 mA, respectively. For synthesis of films with similar thicknesses at lower sputter powers than 300 W, we applied longer deposition times of 1576, 811, 560, 432, and 352 s in order to achieve a constant sputter current-time product at 50, 100, 150, 200, and 250 W, respectively. As substrates, we used pieces cut from an $Al_2O_3(0001)$ wafer with the size 1.25 x 1.25 cm to study epitaxial growth and two 1.25 x 1.25 cm pieces of 1000 Å $SiO_2/Si(100)$ in each run. The sapphire substrate was mounted in the middle of the substrate holder and with the oxidized silicon substrates in positions on each side adjacent to the central position without substrate rotation. The latter was used for thickness measurements only. Prior to deposition, the substrates were degreased in 5 min sequential ultrasonic baths of trichloroethylene, acetone and isopropanol, and blown dry with pure nitrogen. This was followed by an in-situ heat treatment for 1h at 900 °C in the growth chamber before adjusting the temperature to the desired value.

### B. Thin Film Characterization

The films deposited on $Al_2O_3(0001)$ and the $Ti_3SiC_2$ compound target were investigated by X-ray photoelectron spectroscopy (XPS) to determine the chemical bonding structure as well as the elemental composition, using an AXIS Ultra-DLD instrument from Kratos Analytical employing monochromatic Al *Kα* radiation. In order to compensate for charge-up effects due to the insulating properties of the $Al_2O_3(0001)$ substrate, the samples were irradiated by low energy electrons from a flood-gun during the entire analysis. To remove adsorbed contaminants following exposure to the air, the samples were sputter-cleaned for 180 s with 4 keV $Ar^+$ ions incident at a take-off angle of 20°. In the case of the $Ti_3SiC_2$ compound target a longer etching time of 720 s was used to account for the significantly rougher surface. The binding energy (BE) scale was calibrated by setting the Ti $2p_{3/2}$ peak to 454.7 eV corresponding to Ti-C bonding in the analyzed $Ti_3SiC_2$ compound target.[34,35] Quantitative analysis was performed using the Casa XPS software.[36]

The qualitative and quantitative analyses performed by XPS were corroborated by time-of-flight energy elastic recoil detection analysis (ToF-E ERDA) of the films deposited on $Al_2O_3(0001)$. The measurements were carried out with a 36 MeV $^{127}I^{8+}$ ion beam using the set-up at Uppsala University.[37,38] The incident angle of primary ions and exit angle of recoils were both 67.5° to the sample surface normal giving a recoil angle of 45°. The measured ToF-E ERDA spectra were converted into relative atomic concentration profiles using the CONTES code.[39]

The phase composition in the films and the sputtering target were assessed by X-ray diffraction (XRD), conducting θ/2θ scans in a Philips powder diffractometer (PW 1820) with Cu $K\alpha$ radiation at 40 kV and 40 mA. For the applied $Ti_3SiC_2$ target material, the mass fraction of each phase was determined using reference intensity ratio values and scale factors using the Highscore software from Panalytical[40] and the ICDD PDF database.

The cross-sectional transmission electron microscopy (HRTEM) specimens were prepared by gluing the film face to face to form a sandwich specimen, polishing the specimen from both sides to 50 μm, following by ion milling until electron transparency. The details of cross-sectional HRTEM specimen preparation can be found in our previous work.[41] Z-contrast scanning electron microscopy (STEM) together with X-ray energy dispersive spectroscopy (EDX) was carried out in a double Cs-corrected FEI Titan3 60-300 operated at 300 kV, and equipped with a Super-X EDX detector.

The thicknesses of the films deposited on 1000 Å $SiO_2/Si(100)$ substrates were investigated using cross section scanning electron microscopy (SEM, LEO 1550 Gemini) and with images collected at an acceleration voltage of 10 kV.

Room temperature four-point probe measurements were performed on films synthesized on $Al_2O_3(0001)$ with an Auto map system Model 280C from Four Dimensions, Inc. The in-plane resistivity was then calculated by multiplying the obtained sheet resistance with the film thickness obtained by SEM imaging of cross-sections of films deposited on the 1000 Å $SiO_2/Si(100)$ substrates.

## III. RESULTS AND DISCUSSION

Figure 1 is a plot of elemental compositions obtained from ERDA measurements for Ti-Si-C films deposited on $Al_2O_3(0001)$ substrates at temperatures (1a, left) of RT, 300, 500, 600, 700, 800, 850, 900, and 970 ºC and (1b, right) at sputtering powers of 50, 100, 150, 200, 250, and 300 W. The ideal composition of the $Ti_3SiC_2$ compound target with 50 at% Ti, 16.7 at% Si, and 33.3 at% C is indicated as dotted lines. The figure shows that all investigated films exhibit compositions that differ from that of the Ti$_3$SiC$_2$ compound target, *i.e.*, ideally 50 at% Ti, 16.7 at% Si and 33.3 at% C, as indicated by the dotted lines for Ti, Si and C. The ERDA results show that all films exhibit a C content higher than that of the Ti$_3$SiC$_2$ compound target, with 38.5 at% in the film deposited at RT and 47.3 at% when applying 50 W sputtering power and at a substrate temperature of 850 °C. A higher carbon content than 33 at% agrees with studies by Eklund *et al.* for films deposited from a Ti$_3$SiC$_2$ compound target at 4 mTorr and 300 °C[24] as well as 850°C.[25] Alami *et al.*[27] reported carbon deficient films grown with HiPIMS at 90° inclination angle for the growth flux.

The reason for the excess carbon content is that the angular distribution of the species from the target differs for the different elements, with C being much more focused along the *z* axis and the heavier elements (Ti, Si) spread broader.[25] This played critical role in the present geometry with the substrates mounted in the center of the sample holder.

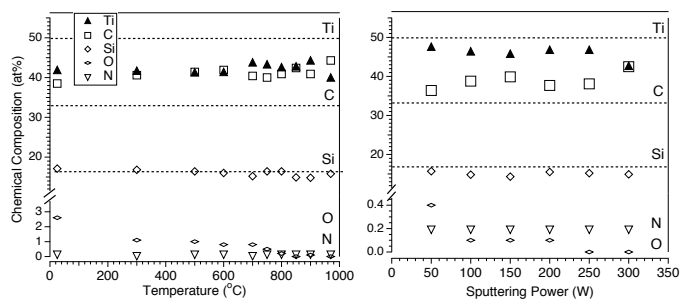

**Figure 1ab.** ERDA compositions for Ti-Si-C films deposited on Al$_2$O$_3$(0001) substrates at temperatures (1a, left) of RT, 300, 500, 600, 700, 800, 850, 900 and 970 °C and (1b, right) at sputtering powers of 50, 100, 150, 200, 250, and 300 W. The ideal composition of the Ti$_3$SiC$_2$ compound target with 50 at% Ti, 16.7 at% Si, and 33.3 at% C is indicated as dotted lines.

Furthermore, for all the investigated films, the Si content is in the region ~15-17 at% *i.e.*, close to that of an ideal target composition. The O content is low, and decreases as the substrate temperature increases, seen from ~2.6 at% in the film deposited at RT towards the detection limit of ERDA. As the growth was carried out at UHV conditions, we suggest that the oxygen originate from the Ti$_3$SiC$_2$ compound target as XPS quantitative analysis show ~8 at% O in the bulk of the material. Schneider *et al.*[42] also suggested the sputtering target as a source of oxygen during sputtering of Cr$_2$AlC films from a Cr$_2$AlC compound target with reported values of ~3 at% at 650 °C.

Figures 2a and 2b show XPS measurements at the Ti *2p$_{3/2,1/2}$*, C *1s*, Si *2s*, and Si *2p* edges of Ti-Si-C films deposited with a sputtering power of 50 W or 300 W at 850 °C or at RT at a sputtering power of 300 W, in comparison to the Ti$_3$SiC$_2$ compound target. The bulk material has photoelectron peaks at 454.7 eV, 460.6 eV, 282.0 eV, 150.4 eV, and 99.1 eV binding energies, respectively. These values are in good agreement with XPS measurements of a reactively hot-pressed Ti$_3$SiC$_2$ sample by Stoltz *et al.*[43] with 454.74 eV for Ti *2p$_{3/2}$*, 460.8 eV for Ti *2p$_{1/2}$*, 281.83 eV for C *1s*, 98.91 eV for Si *2p$_{3/2}$* and 99.52 eV for Si *2p$_{1/2}$*.

As the investigated films were deposited on Al$_2$O$_3$(0001) substrates, charge compensation was necessary by means of setting the Ti *2p$_{3/2}$* peak in Fig. 2a (top panel) to 454.7 eV *i.e.*, the same binding energy as recorded for the bulk Ti$_3$SiC$_2$ compound target and corresponding to Ti-C bonding[35,44] marked with a dotted line. From the Ti *2p$_{3/2,1/2}$* peak structures, we observe that the film XPS peaks align to those of the Ti$_3$SiC$_2$ compound target, where all the investigated samples have a spin-orbit splitting of 5.9 eV. The energy calibration of the Ti *2p$_{3/2,1/2}$* peaks agrees with the study by Eklund *et* *al.*,[34] investigating epitaxial TiC(111) and Ti$_3$SiC$_2$(0001) films, comparing the chemical bonding structure to nc-TiC/a-SiC films.

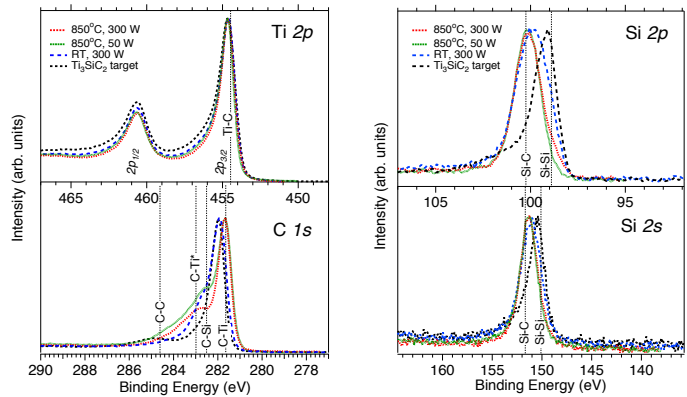

**Figure 2ab.** XPS Ti *2p*, C *1s*, Si *2s* and Si *2p* spectra recorded from the Ti$_3$SiC$_2$ compound target after 720 s of sputter-cleaning and Ti-Si-C films deposited on Al$_2$O$_3$(0001) substrates at temperatures of 850 °C and with sputtering powers of 50 or 300 W, or at RT and a sputtering power of 50 W following 180 s of sputter-cleaning. Litterature binding energies for Ti-C,[44] C-Ti,[35] C-Si,[47] C-Ti*,[48] C-C,[48] Si-C,[47] and Si-Si[43] are indicated by the verical dotted lines.

The bottom panel in Fig. 2a shows C *1s* photoelectron peaks of the films deposited with 50 and 300 W sputtering power that are located at a lower binding energy of 281.7 eV compared to the Ti$_3$SiC$_2$ compound target and the RT film which both show 282.0 eV. The C-Ti peak at 281.7 eV coincides with values between 281.6 -281.8 eV[35,45,46] as indicated by a dotted vertical line in Fig. 2a, bottom panel. When compared to the Ti$_3$SiC$_2$ compound target, the C *1s* peaks of the films have more pronounced tails towards higher binding energy, with literature values of C-Si bonding at 282.5 eV,[47] C-Ti* at 283.0 eV,[48] as indicated by a dotted vertical lines in Fig. 2a, bottom panel. A closer inspection of the intensities in the tails of the C *1s* data suggests more pronounced C-Si bonding in the film deposited at 850 °C with 50 W, the film with the highest carbon content of 47.3 at%, whereas the C-Ti* contribution appears to increase when increasing the sputtering power to 300 W. These observations are correlated with a possible interface state or phase contribution as previously observed as a feature due to charge-transfer, denoted C-Ti$^*$ in C 1*s* XPS spectra from nanocystalline-TiC in an amorphous C matrix[48,49] also interpreted as sputtering damage.[50] From XPS, the tail from the film deposited at RT and 300 W sputtering power shows contribution both from C-Si and C-Ti* in this film with a C content of 38.5 at%. In addition, there is a signature of C-C bonds as indicated by a dotted vertical line in Fig. 2a, bottom panel at 284.6 eV,[48] most pronounced for the Ti$_3$SiC$_2$ target.

In Fig. 2b (top panel), the binding energies of the Si *2p* peaks are higher in the films seen from: 99.85 eV for the film deposited with 300 W sputtering power at RT condition, 100.15 eV for the film deposited with 50 W sputtering power at 850 °C, and 100.15 eV for the film deposited with 300 W sputtering power at 850 °C compared to that of the Ti$_3$SiC$_2$ compound target at 99.1 eV.

The Si *2p* peaks of the films deposited with 50 and 300 W sputtering power are close to the reported value of the Si-C bonding at 100.3 eV[47] (indicated by a dotted vertical line in Fig. 2b, bottom panel) and Si-C bonding is supported by the C *1s* XPS region. The film grown without external heating shows a somewhat lower binding energy of 99.9 eV *i.e.*, closer to Si-C bonding than Si-Si bonding at 98.91 eV.[43]

The observations made from the Si *2p* peak positions in Fig. 2b are consistent with the peak positions of the Si *2s* spectra in the bottom panel, *i.e.*, the Si *2s* binding energies are higher than in the Ti$_3$SiC$_2$ compound target at 150.4 eV compared to the film peaks at 151.05 eV (RT), 151.35 eV (50 W) and 151.35 eV (300 W). For comparison, the literature value for the Si-Si bonding is 150.50 eV[51] and 151.70 eV for Si-C.[52]

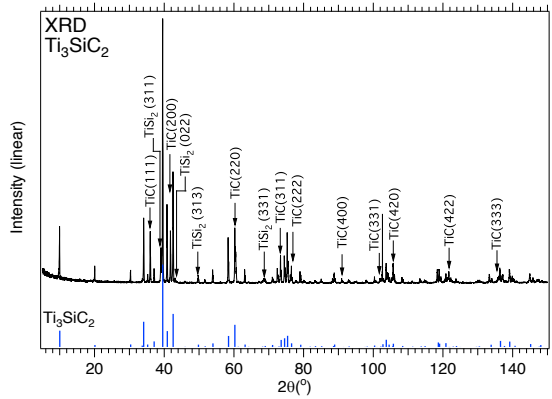

**Figure 3.** X-ray θ/2θ scan recorded from the for thin film growth applied $Ti_3SiC_2$ compound target. The arrows indicate reflections from the TiC and $TiSi_2$ minority phases. The bars below the diffractogram represent peak positions and intensities from $Ti_3SiC_2$ in reference.[53]

Figure 3 shows an X-ray θ/2θ scan recorded from the $Ti_3SiC_2$ compound target. The diffraction pattern displays all prominent $Ti_3SiC_2$ peaks listed in the reference diffraction pattern,[53] and where the intensity distribution among the $Ti_3SiC_2$ peaks supports a randomly oriented target material.[53] $Ti_3SiC_2$ reflections from the reference diffraction pattern are indicated by the vertical bars at the bottom of Fig. 3. In addition, the diffractogram shows clearly visible peaks from TiC 111, 200, 220, 311, 222, 400, 331, 420, 422, and 511/333 seen from increasing 2θ angles.[54]

Peaks at 2θ angles of 39.1°, 43.2°, 49.7°, and, 68.7° correspond to the 311, 022, 313, and 331 peaks in orthorhombic $TiSi_2$, respectively.[55] On the other hand, $Ti_5Si_3$ cannot be found in the diffraction pattern. Quantitative XRD analysis revealed that the target material consists of only a 78% mass fraction of $Ti_3SiC_2$, while TiC and $TiSi_2$ constitute the remaining 13% and 9%, respectively. The presence of minority phases has also been observed in other target MAX-phase materials such as $Ti_2AlC$[56] and $Cr_2AlC$,[42] but typically at much lower content (< 5%). As previously mentioned, quantitative analysis from XPS spectra yielded an O content of ~7.7 at.% and with a bulk composition of 36.6 at% Ti, 18.7 at% Si, 37.0 at% C *i.e.*, that is different from the ideal 50 at% Ti, 16.7 at% Si and 33.3 at% C. It is likely that the properties determined from the sputtering source will affect the possibilities to deposit epitaxial $Ti_3SiC_2$.

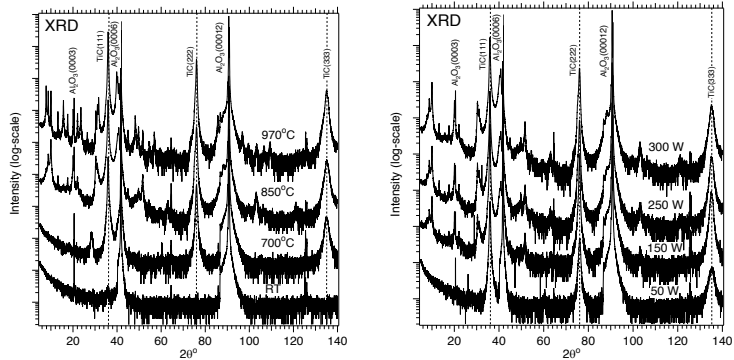

**Figure 4ab.** X-ray θ/2θ scans from Ti-Si-C films deposited on $Al_2O_3$(0001) substrates using substrate temperatures of RT, 700, 850 and 970 °C at a sputtering power of 300 W (4a, left) as well as sputtering powers of 50, 150, 250, and 300 W (4b, right).

Figure 4 shows X-ray θ/2θ scans of Ti-Si-C films deposited on $Al_2O_3$(0001) substrates as functions of temperature (4a, left) and power (4b, right). For growth at RT, the diffraction pattern only displays a weak peak at 2θ≈36° from TiC 111, and where a potential TiC 200 peak overlaps with the $Al_2O_3$ 0006 peak. At 300 and 500 °C, the TiC 111 peak increase in intensity and TiC 220 and TiC 222 peaks appear at 2θ≈60.5° and 2θ≈76.1°, respectively (not shown). At 700 °C, there are clear peaks from TiC 111, 222, and 333, indicating oriented growth.[57] In addition, there is a peak at 2θ≈30°, corresponding to the Nowotny phase $Ti_5Si_3C_x$. Emmerlich *et al.*[18] reported growth of

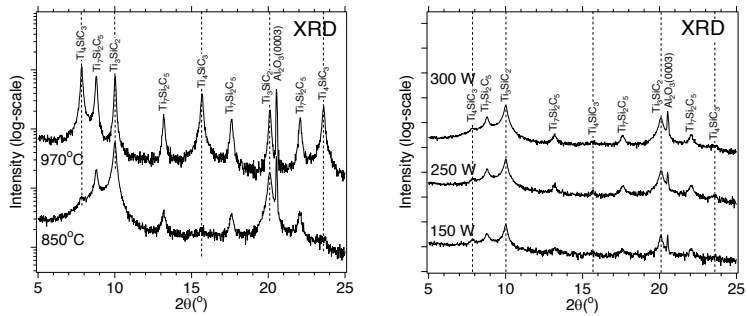

**Figure 5ab.** Diffraction patterns at low diffraction angles for the Ti-Si-C films deposited on Al$_2$O$_3$(0001) substrates at temperatures of 800, 850 and 970 °C and at a sputtering power of 300 W (5a, left) and at sputtering powers of 150, 250, and 300 W at a temperature of 850 °C (5b, right).

Ti$_5$Si$_3$C$_x$ at 700 °C by sputtering from elemental sources, and it has also been deposited by sequential growth by Vishnyakov *et al.*[29] At 850 °C, there are peaks at low angles (for clarity, see Fig. 5a) that originate from Ti$_3$SiC$_2$, Ti$_4$SiC$_3$, and the intergrown Ti$_7$Si$_2$C$_5$ phase. A Ti$_3$SiC$_2$ peak of low-intensity and positioned at 2θ=10.0° is visible already at a deposition temperature of 800 °C. Increasing the temperature to 970 °C results in higher intensities for the peaks from Ti$_3$SiC$_2$, Ti$_4$SiC$_3$ and Ti$_7$Si$_2$C$_5$, which is most pronounced for the Ti$_4$SiC$_3$ phase. This agrees with observations showing that growth of Ti$_4$SiC$_3$ is favored at a temperature of 1000 °C.[58] From the bottom diffractogram in the right panel of Fig. 4b, it is seen that a sputtering power of 50 W at 850 °C results in growth of 111-oriented TiC. Tripling the sputtering power to 150 W results in nucleation of Ti$_3$SiC$_2$, Ti$_4$SiC$_3$ and the intergrown Ti$_7$Si$_2$C$_5$ structure as observed by the peaks in Fig. 5b, right panel. The formation of TiC should be favored by a C-rich composition.[59]

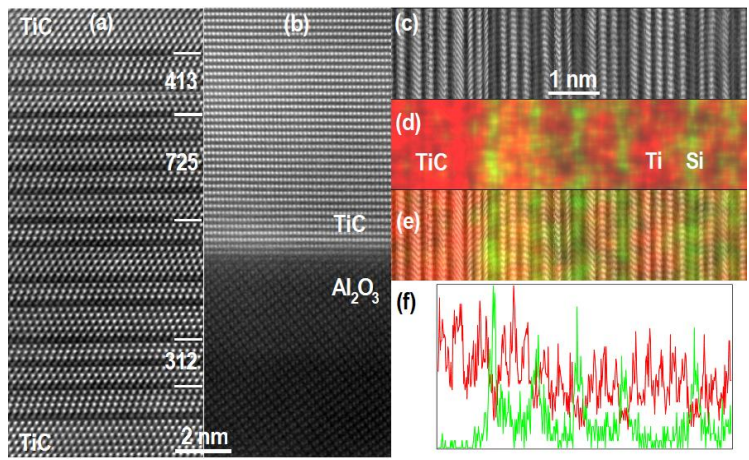

**Figure 6.** (a) shows a HRTEM image of MAX phases Ti$_3$SiC$_2$, Ti$_4$SiC$_3$, and Ti$_7$Si$_2$C$_5$ embedded in TiC, (b) STEM image of an interface between the TiC film and Al$_2$O$_3$ substrate, (c) EDX mapping of the STEM image, (d) Ti+Si map, where red/dark represents Ti and and green/light represents Si, (e) the STEM image superimposed with Ti+Si map (Ti in red/dark and Si in green/light), (f) the corresponding line scan along 0001 direction (Ti in red/dark and Si in green/light).

Figure 6a shows a HRTEM image that illustrates intergrowth of the MAX phases Ti$_4$SiC$_3$, Ti$_5$Si$_2$C$_5$, and Ti$_3$SiC$_2$ between TiC grains. This has previously been observed in refs. 17 and 18. The Ti$_3$SiC$_2$, Ti$_4$SiC$_3$, and Ti$_7$Si$_2$C$_5$ phases are consistent with the θ/2θ diffraction patterns in Figs. 4 and 5. An interesting structure consisting of a half unit cell of Ti$_3$SiC$_2$ together with a half lattice of Ti$_5$SiC$_4$ observed between Ti$_7$Si$_2$C$_5$ and Ti$_3$SiC$_2$, indicating a potential new structure of Ti$_8$Si$_2$C$_6$, that appears similar to the Ti$_7$Si$_2$C$_5$ structure. The corresponding STEM image gives the epitaxial relationship between the MAX phases and the TiC matrix, that is: (0002)[1$\bar{1}$00]MAX//(111)[11$\bar{2}$]TiC. The MAX phases appear to nucleate and grow inside of the TiC. From HRTEM, we find no MAX-phase nucleation directly formed on the surface of Al$_2$O$_3$ substrate, which is different from previous studies on Ti$_3$AlC$_2$[60-63]. Instead, TiC is nucleated on the Al$_2$O$_3$ substrate with an epitaxial relationship of (111)[11$\bar{2}$]TiC//(0002)[1$\bar{1}$00]Al$_2$O$_3$ as shown in Fig. 6b. Twinning of the TiC is known to favor the nucleation of MAX-phases by forming trigonal prismatic sites for the larger silicon atoms to accommodate in the structure[64], which was frequently observed in the film. Twinning often appears as Moire fringes with plane distances of three times of that of the TiC (111). Here, care was taken to distinguish the MAX-layered structures from the Moire fringes originating from TiC(111) in the HRTEM images. To verify the variation of the chemical composition in the MAX phases, EDX mapping was carried out and the results are shown in Fig. 6c-f. The composition was obtained from EDX analysis of the Ti$_3$SiC$_2$-phase yielded a ratio of Ti:Si about 4:1. Thus, the Si content is lower than nominal for the Ti$_3$SiC$_2$, which is an indication of possible vacancies in the Si layers.

The formation of an epitaxial TiC "incubation layer" typically occurs for epitaxial growth of MAX phases. High supersaturation of the A element is usually needed before the elements to partition to form the MAX phases.[17,18,25] Furthermore, the TEM micrograph in Fig. 6a shows that TiC is heavily twinned with either ABC or BCA stacking and where the twining yields a second in-plane relationship of TiC[0$\bar{1}$1]//Al$_2$O$_3$[10$\bar{1}$0]. This has been observed by Palmquist *et al.*[17] and Emmerlich *et al.*,[18] and was explained by Si lowering the twin-fault energy for TiC(111).

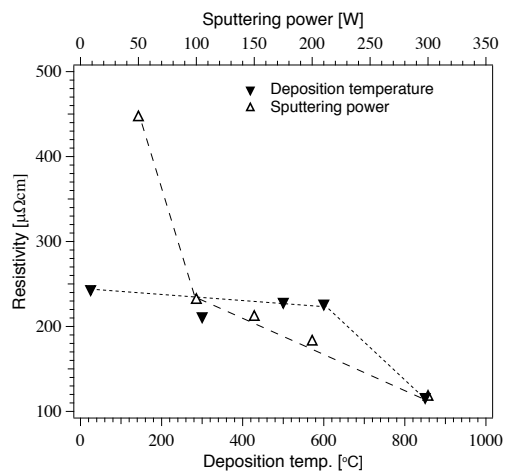

**Figure 7.** Four-point probe resistivity of Ti-Si-C films deposited on Al$_2$O$_3$(0001) substrates using substrate temperatures of RT, 300, 500, 600 and 850 °C at a sputtering power of 300 W as well as sputtering powers of 50, 100, 150, 200 and 300 W at a substrate temperature of 850 °C. The dashed lines are trend guides for the eye.

Figure 7 shows four-point probe resistivity measurements on films deposited on Al$_2$O$_3$(0001) substrates using substrate temperatures ranging between RT to 850 °C at a sputtering power of 300 W as well as sputtering powers between 50 and 300 W at substrate temperatures of 850 °C. As observed, the measured resistivity values are in the range ~120 to ~450 μΩcm and with the lowest resistivity in the film deposited at 850 °C, using a sputtering power of 300 W. These values are an order of magnitude higher than measured for epitaxial Ti$_3$SiC$_2$ films in the region of 20-25 μΩcm.[17] We attribute this to less-conductive TiC inclusions as supported by the TiC peaks of high intensity in the θ/2θ diffractograms and the low-resolution TEM image in Fig. 6a; the measured resistivity values for epitaxial TiC films being in the region of 200-260 μΩcm.[17]

From the ERDA results, it is clear that the stoichiometry of the Ti$_3$SiC$_2$ compound target is not preserved in the deposited Ti-Si-C films. This stoichiometry difference has been reported before,[24-27] and occurs because of the deposition geometry and difference in angular distribution of the elements, as discussed earlier. The fact that the films are C-rich seems to favor the growth of TiC, which is possibly due to: (i) the strong driving force of the metal Ti to form carbides and (ii) the large homogeneity range of the phase, ranging from $TiC_{0.47}$ to $TiC_{0.99}$,[59] to compare with 33 at% $Ti_3SiC_2$, ~35.7 at% $Ti_7Si_2C_5$ and 37.5 at% $Ti_4SiC_3$. The diffraction patterns in Fig. 4 indicate from the 2θ angles of the position of the peaks a TiC 111, 222, 333, a C-rich composition for the deposited TiC.[54] Thus, the excess C favor to form TiC instead of the $Ti_3SiC_2$, $Ti_7Si_2C_5$, and $Ti_4SiC_3$ phases. Growth at high temperature ≥ 850 °C, favors growth of $Ti_3SiC_2$, $Ti_7Si_2C_5$, $Ti_4SiC_3$ phases as shown by the XRD and the TEM. Increasing the temperature to 970 °C favors the growth of the $Ti_4SiC_3$ phase, with a higher C content compared to $Ti_3SiC_2$ seen from 37.5 at% C in 413 and 33 at% in 312. Interestingly, the Si-content is lower than in $Ti_3SiC_2$ seen from seen from 12.5 at% in $Ti_4SiC_3$ and ~16.7 at% in $Ti_3SiC_2$.

This occurs because of the higher vapor pressure of Si compared to the other elements, resulting in a reduced sticking coefficient of Si. The evaporation of A-elements is a general phenomenon in MAX-phase growth and even more pronounced for elements with higher vapor pressures (*e.g.*, Al and In).[65] This implies that growth of $Ti_3SiC_2$ from a compound source is restricted to a certain temperature range to avoid Si evaporation and the on-set for growth of phases with a lower Si content such as $Ti_4SiC_3$ and $Ti_7Si_2C_5$ with 12.5 at% and 14.3 at% Si, respectively.

Our results show the difficulties in depositing single-phase and epitaxial $Ti_3SiC_2$ films from a $Ti_3SiC_2$ compound target. From XRD, we find TiC, $Ti_4SiC_3$ and $Ti_7Si_2C_5$ as competing phases and where growth of $Ti_3SiC_2$ seems restricted by temperature below 1000 °C. The difficulties in controlling the processes is mainly attributed to the angular dependence of the sputtered species from the target at the applied process conditions. However, the properties determined for the applied $Ti_3SiC_2$ target seen from minority phases and oxygen contaminants set further restrictions on the epitaxial thin film growth. For this, we acknowledge the development of sputtering targets with improved properties as exemplified for growth of $Ti_2AlC$[66,67] and $Ti_2AlN$,[68,69] respectively.

## IV. CONCLUSIONS

From magnetron sputter deposition of a $Ti_3SiC_2$ compound target, we show that epitaxial TiC, $Ti_3SiC_2$, $Ti_4SiC_3$, and $Ti_7Si_2C_5$ can be deposited from a $Ti_3SiC_2$ compound target at temperatures of 850 °C and above. Higher sputtering powers applied to the target favors epitaxial growth of $Ti_{n+1}SiC_n$ phases and yields films of lower resistivity values. The composition of the grown films has a higher carbon content than in the $Ti_3SiC_2$ phase due to differences in the angular distribution of C, Si, and Ti during sputtering. XRD of the $Ti_3SiC_2$ compound target shows that it contains TiC and $TiSi_2$ as minority phases as well as ~8 at.% O according to XPS. Thus, growth of epitaxial singe-phase $Ti_3SiC_2$ films requires a growth flux with strict 3Ti:Si:2C composition as well as a temperature in the region of 800 °C.

## ACKNOWLEDGMENTS

We acknowledge funding from the Swedish Government Strategic Research Area in Materials Science on Functional Materials at Linköping University (Faculty Grant SFO-Mat-LiU No. 2009-00971). MM acknowledges financial support from the Swedish Energy Research (no. 43606-1) and the Carl Tryggers Foundation (CTS16:303, CTS14:310). PE acknowledges the Knut and Alice Wallenberg Foundation through the Wallenberg Academy Fellows program. GG acknowledges financial support from the Åforsk Foundation Grant 16-359, and Carl Tryggers Foundation (CTS 17:166). The authors acknowledge Åke Öberg at ABB Sverige AB for the target material and, Uppsala University for access to the Tandem Laboratory.